\documentstyle[aps,preprint,epsf]{revtex}

\def\beq{\begin{equation}}
\def\eeq{\end{equation}}
\def\ie{i.e.}
\def\etal{{\it et al. }}
\def\prb{Phys. Rev. B }
\def\prl{Phys. Rev. Lett. }
\def\mpl{Mod. Phys. Lett. B }

\begin{document}

\draft
\title{Comparison between the two models of dephasing in mesoscopic systems}
\author{T. P. Pareek \cite{tpp}, Sandeep K. Joshi \cite{skj} and  A. M. 
Jayannavar \cite{amje}}
\address{ Institute of Physics, Sachivalaya Marg, Bhubaneswar 751 005, India}

\maketitle

\begin{abstract}

In mesoscopic systems to study the role of inelastic scattering on the
phase coherent motion of electrons two phenomenological models have been
proposed. In the first one, due to B\"uttiker, one adds a voltage probe
into the system (or in the scattering matrix). The second model invokes 
the complex (or optical) potential in the system Hamiltonian. Studying 
a simple geometry of a metallic loop in the presence of Aharonov-Bohm 
magnetic flux, we show that the two probe conductance is symmetric in the 
reversal of the magnetic field in B\"uttiker's approach. Whereas the two 
probe conductance within the complex potential model is asymmetric in 
the magnetic flux reversal contrary to the expected behavior.  

\end{abstract}

During the last two decades, the study of transport in mesoscopic systems
has been actively pursued owing to immense interest from technological as
well as from fundamental view point \cite{kramersNato,altshular,vonhouten,beaumont,ando,datta,washburn,ferry}. 
Mesoscopic systems are structures
made of metallic or semiconducting material on a nanometer scale. The
length scale associated with the dimensions in these systems is
much smaller than the inelastic mean free path or the phase breaking length
$(L_\phi)$. The phase breaking length (or phase coherence length),
$L_\phi$, is the average diffusion length between the two inelastic
collisions. Typically $L_\phi$ scales with the temperature $T$ in a power
law form, \ie, $L_\phi~=~T^{-p}$ ($p$ lies in the range 1 and 2). At low
enough temperature when the system size $L$, is much smaller than the phase
breaking length $L_\phi$, an electron maintains phase coherence across the
entire sample. The mesoscopic sample should be treated as a quantum
scatterer. Here shape of the sample, quantization of energy levels and
discreteness of charge play a major role. Thus mesoscopic systems have 
provided an opportunity of exploring truly quantum mechanical effects 
beyond atomic realm. In the quantum-phase coherent
transport regime classical Ohm's law breaks down \cite{datta} in the sense that if one
adds two resistors having resistance $R_1$ and $R_2$ in series then the
total resistance $R$ of the system is no longer a sum of the two
resistances $R_1$ and $R_2$ ($R~\neq~R_1+R_2$).  Apart from this in lower
dimensions resistance is a non-self averaging quantity in that
the resistance fluctuations over the ensemble of macroscopically identical
samples dominates the ensemble average \cite{datta,fuku,nkuamj}. The quantum resistance of a sample
depends on the details of the relative position of scatterers.
Thus the mesoscopic system is characterized by the sample
specific global resistance. However, as the temperature increases, inelastic
scattering effects start dominating leading to the loss of phase coherence.
If the sample size is larger than $L_{\phi}$ the sample breaks
up dynamically into mutually incoherent domains of size $L_{\phi}$, with
transport within each domain remaining phase coherent. Here self-averaging
property of the resistance is automatically realized and 
classical additivity of resistance is restored \ie, 
$R~=~R_1+R_2$. In the phase coherent transport regime several, often
counter-intuitive, new experimental results have been obtained \cite{kramersNato,altshular,vonhouten,beaumont,ando,datta,washburn,ferry} and have
been successfully explained within a Landauer-B\"uttiker formalism for dc
transport \cite{rolf1,butt4p}. 

Although experiments on mesoscopic samples confirm the predictions based
on the phase coherent transport theory, a quantitative comparison at finite
temperatures requires the loss of phase coherence be included in to the
theory. There are two widely used phenomenological models which have been
proposed for this purpose. In the first method due to B\"uttiker, one
introduces a fictitious voltage probe in the scattering matrix \cite{datta,buttibm,buttvp}. 
The voltage probe breaks the phase coherence by removing electrons from the
phase coherent motion in the mesoscopic system and subsequently
reinjecting them without any phase relationship. The treatment based on
the voltage probe method (which serves as a inelastic scatterer) has been
extended to include the realistic physics of inelastic processes
occurring uniformly in space \cite{datta}. To simulate inelastic scattering other
method makes use of complex (or optical) potentials \cite{lee,zohta,beenak}. In that case the
Hamiltonian becomes non-Hermitian and thus the particle number is not
conserved. In these studies the absorption is identified as the spectral
weight lost in the inelastic channels. As an example, in the case of
one-dimensional double barrier structures the absorbed part is assumed to 
tunnel through both the left and the right hand sides of the barriers in
proportion to the transmission coefficient of each barrier, and this is
added to the coherent transmission to get the overall transmission
coefficient \cite{zohta}. It should be noted that in the presence of imaginary
potentials the temporal coherence of the wave is preserved in spite of
absorption which causes a particle non-conserving scattering process. The
absorption is to be understood as a depletion of the coherent amplitude by
the inelastic process. Problems related to the use of complex potentials have
been discussed in the earlier literature \cite{rubio,amj,gupta}. A recent
study identifies the limit in which these two models of dephasing are 
equivalent and the distribution of conductance in that limit has been 
calculated \cite{beenak}.  

In our present study we analyse both these models in the presence of
magnetic flux, and show that these two models lead to qualitatively
different results for the symmetry of the two probe conductance in the
presence of magnetic field. In the B\"uttiker's approach of voltage probe,
two probe conductance is symmetric in the reversal of magnetic field as
has been observed experimentally \cite{datta}. However, the model based on the complex
potential makes the two probe conductance asymmetric in the magnetic field
reversal contrary to the expected behavior \cite{datta}. 

\begin{figure}
\protect\centerline{\epsfxsize=4in \epsfbox{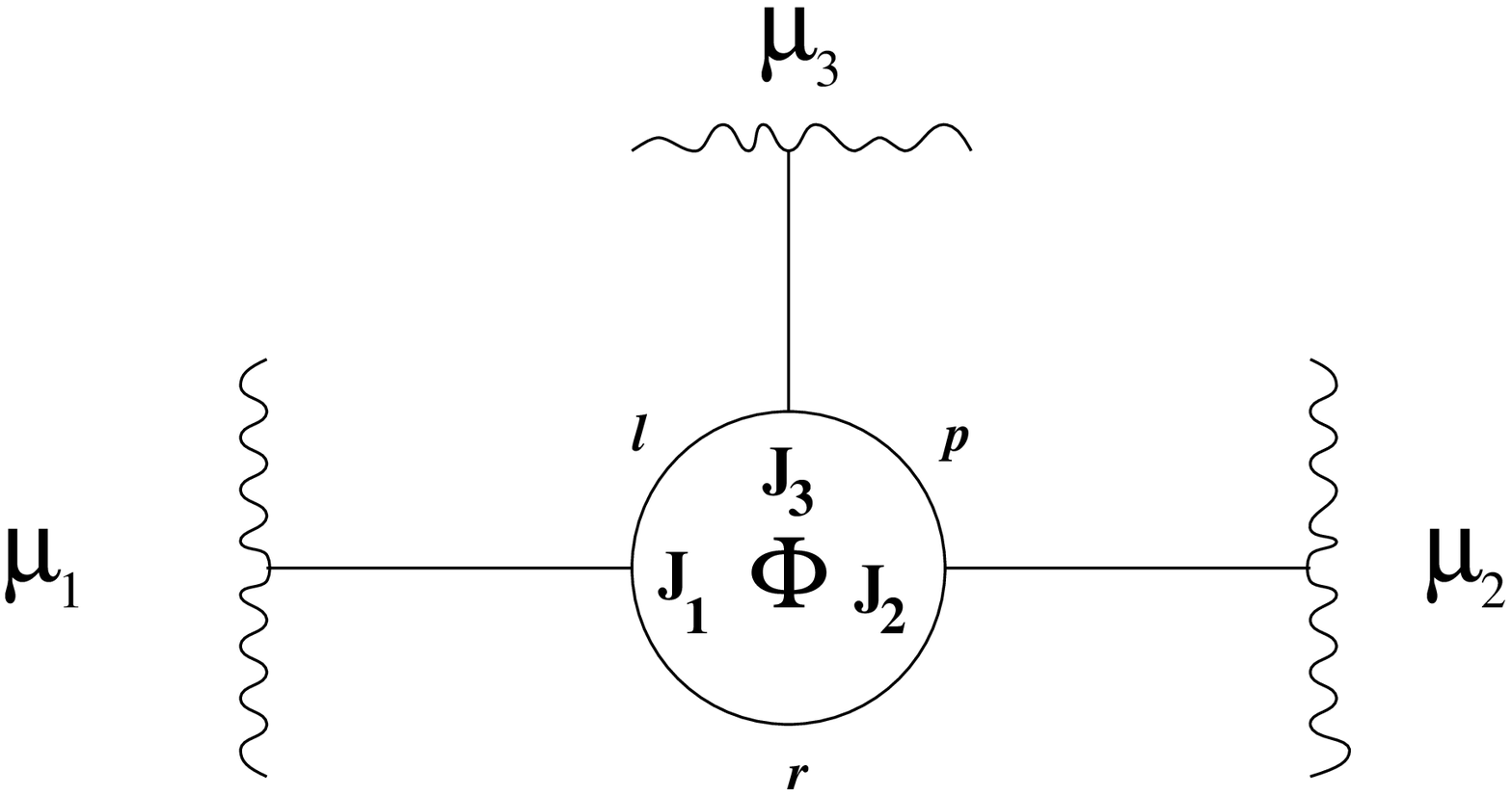}}
\vspace{0.5in}
\caption{A metallic loop connected to three reservoirs in the presence of 
magnetic flux $\Phi$.}
\label{res}
\end{figure}

\begin{figure}
\protect\centerline{\epsfxsize=5in \epsfbox{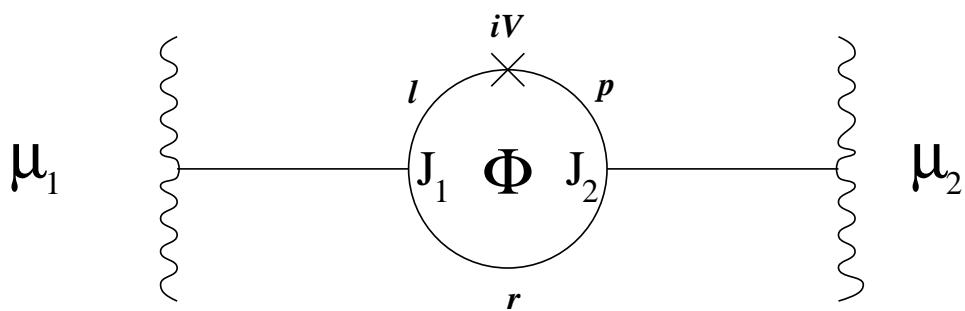}}
\vspace{0.5in}
\caption{A metallic loop connected to two reservoirs in the presence of a 
magnetic flux $\Phi$ and a $\delta$-function imaginary potential $iV$.}
\label{iv}
\end{figure}

To this end we consider a simple geometry of a one-dimensional metallic ring
in the presence of Aharonov-Bohm (A-B) flux as shown in Fig. \ref{res} and 
Fig. \ref{iv}. In Fig. \ref{res} we have attached an additional lead at 
point x on the upper arm of the loop which acts as a voltage probe. In Fig.
\ref{iv} we have introduced a $\delta$-function optical potential of
strength $iV$ at the same point x in the upper arm which acts as an
absorber. The length of the upper arm is $l+p$ and that of the lower arm 
is $r$. The total circumference of the loop is $L=l+p+r$.


The two probe phase coherent conductance of a mesoscopic sample at zero
temperature is given by the Landauer formula \cite{rolf1}. 
\beq
\label{LandauerF}
G ~ = ~ \left ( \frac{e^2}{h} \right ) ~ T, 
\eeq where
$T$ is the transmission probability for carriers to traverse the sample.
Here the transmission probability is taken at the Fermi energy.  From
Eqn. \ref{LandauerF} we introduce a dimensionless conductance
$g~=~(h/e^2)~G~=~T$. 

To study the effect of dephasing in the presence of A-B flux via 
B\"uttiker's approach we consider a mesoscopic open
ring connected to three electron reservoirs at chemical potential $\mu_1$,
$\mu_2$ and $\mu_3$ as shown in Fig. \ref{res}. An A-B flux $\phi$ is
present at the center. We focus on the situation when third lead is used
as a voltage probe to measure the chemical potential $\mu_3$. The net
current in third lead is zero. If we denote transmission probabilities of
carriers incident in lead $j$ to reach lead $i$ by $T_{ij}$
($i,j~=~1,2,3$), then the two probe conductance ( in the dimensionless
units ) of the A-B ring is given by \cite{datta},
\beq
\label{ButtF}
g_B~=~T_{21}~+~\frac{T_{31}T_{23}}{T_{31}+T_{32}}  
\eeq
We see that the two port conductance is a sum of two parts. The first part
$T_{21}$ arises due to those electrons which traverse the ring without
ever entering into the third reservoir, this corresponds to the elastic
transmission probability. The second part \ie, $
T_{31}T_{23}/(T_{31}+T_{32}) $ describes electrons which emanate from port
1, reach reservoir 3 where their energy and phases are randomized and from
reservoir 3 in an additional step reach reservoir 2. In this sense the
third lead connected to reservoir acts like an inelastic scatterer. 

In the presence of absorbing potential ( Fig. \ref{iv} ) the sum of
transmission ( $T$ ) and reflection coefficient ( $R$ ) is not unity. The
absorption coefficient is given by $A~=~1-T-R$. In this model of dephasing
absorbed part is assumed to be re-emitted to the right and left in
proportion to the transmission coefficient at the right and left hand side
of the absorber. In our case absorbed flux of particles is re-emitted
equally on both sides of the absorber and consequently the dimensionless
conductance in this model is given by 
\beq 
\label{OptF} g_i~=~T~+~A/2 
\eeq
To calculate conductances $g$, $g_B$ and $g_i$ we need to know
transmission and reflection coefficients. To calculate them we follow our
earlier method of quantum waveguide transport on networks 
\cite{amjpsd,psdamj,tpppsdamj,tppamj}.  Our
calculation is for a non-interacting system of electrons. We set units of
$h$, $e$ and $m$ to be unity. We do not assume
any particular form for the scattering matrix for the junctions $J_1$,
$J_2$ and $J_3$, but scattering at junctions follows from the first
principles using quantum mechanics. We have imposed Griffith's boundary
conditions (conservation of current) and single valuedness of the
wavefunctions at the junctions. After calculating different transmission
and reflection coefficients we substitute them back into Eqns. 
(\ref{ButtF}) and (\ref{OptF}) to get the analytical expressions for the
conductances $g_B$ and $g_i$.  However, the analytical expressions are too
long to be reproduced here. In the following we present our results
graphically. 

\begin{figure}
\protect\centerline{\epsfxsize=5in \epsfbox{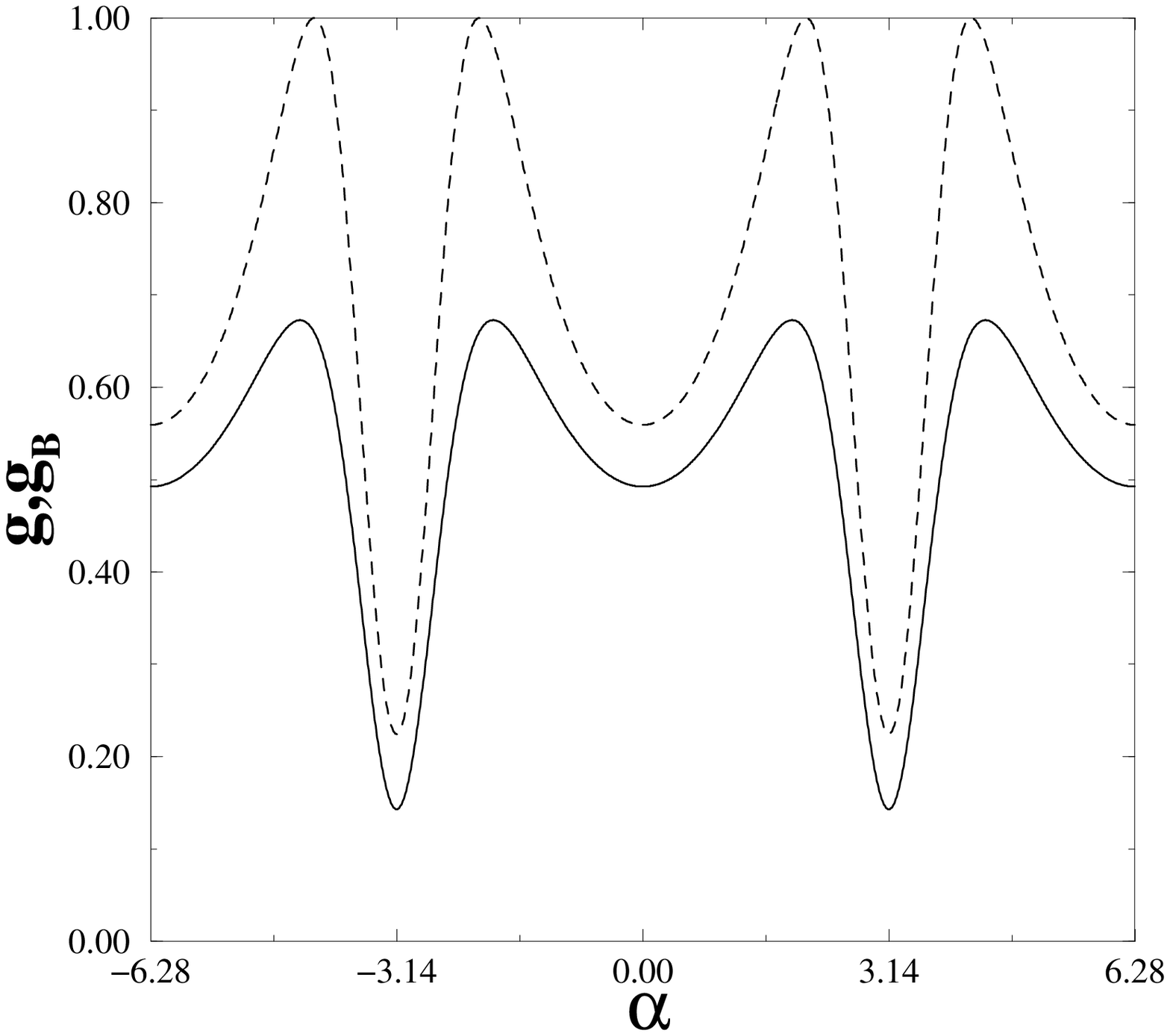}}
\vspace{0.5in}
\caption{The plots of $g$ (dotted line) and $g_B$ (solid line) versus $\alpha~
=~2\pi\phi/\phi_0$}
\label{Fgb}
\end{figure}

In Fig. \ref{Fgb} we plot the dimensionless conductance $g$ (dotted line) 
and $g_B$ (solid line) as a function of the dimensionless flux 
$\alpha~=~2\pi\phi/\phi_0$, where $\phi_0~=~hc/e$ being the elementary 
flux quantum. We choose $kL~=~5$, $l/L~=~0.15$, $p/L~=~0.3$ and 
$r/L~=~0.55$. Both $g$ and $g_B$ oscillate with a period $\phi_0$ and are 
symmetric with respect to the field reversal as expected for the two 
probe conductance. As $g_B$ includes the effect of dephasing due to an 
additional voltage probe, the amplitude of oscillations in $g_B$ are 
smaller than that observed for $g$, the phase coherent conductance. This 
is expected from the fact that inelastic effects reduces the amplitude of 
conductance oscillations (or interface effects) \cite{washburn}.

\begin{figure}
\protect\centerline{\epsfxsize=5in \epsfbox{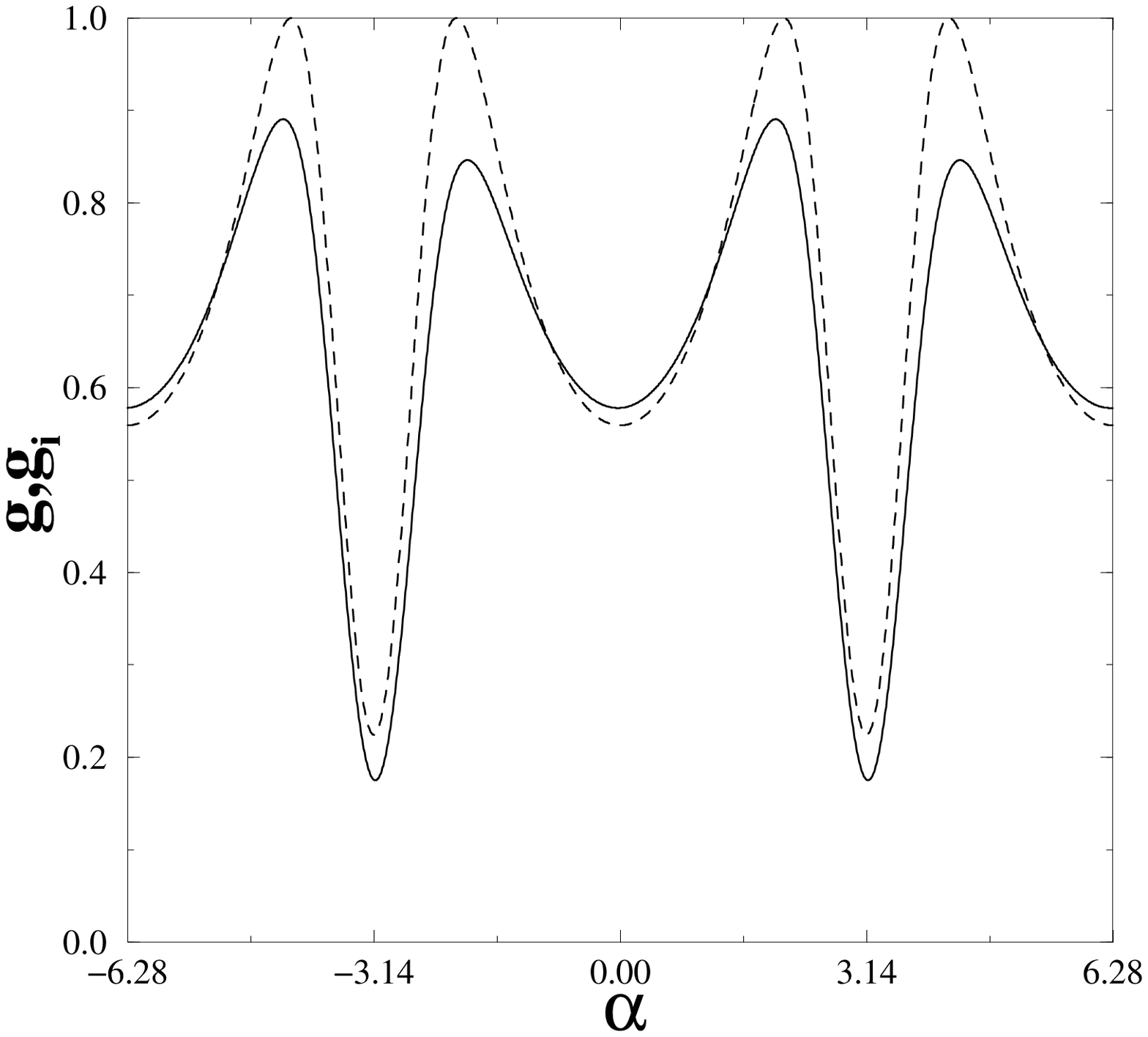}}
\vspace{0.5in}
\caption{The plots of $g$ (dotted line) and $g_i$ (solid line) versus $\alpha~
=~2\pi\phi/\phi_0$}
\label{Fgi}
\end{figure}

In Fig. \ref{Fgi} we plot $g$ (dotted line) and $g_i$ (solid line) as 
a function of $\alpha$. The length parameter values are the same as used 
for Fig. \ref{Fgb}. The strength of the imaginary potential in evaluating 
$g_i$ is taken to be $VL~=~3$. Both $g$ and $g_i$ are periodic in flux 
with a period $\phi_0$. The amplitude of oscillations in $g_i$ are 
smaller than that observed for $g$. However, $g_i$ is asymmetric in 
field reversal in contrast to the expected behavior. This also 
follows from our analysis of the symmetries of the transmission and 
reflection coefficients under the field reversal in the presence of 
complex potential ($iV$), namely, $T(V,\phi)~\neq~T(V,-\phi)$ and 
$R(V,\phi)~=~R(V,-\phi)$.

In conclusion we have compared two phenomenological models for dephasing 
in mesoscopic systems in the presence of Aharonov-Bohm flux. The model 
due to B\"uttiker based on addition of voltage probe to simulate 
inelastic scattering leads to the two probe conductance which is symmetric in 
magnetic field. On the other hand model based on the use of complex 
potential leads to the two probe conductance which is asymmetric in the 
magnetic field contrary to the expectation based on experimental as well 
as theoretical predictions. We would also like to emphasize that the use 
of imaginary potentials is justified in the case of optical wave 
propagation in a absorbing or a lasing medium (random dielectric media). 
In the electromagnetic wave propagation the bosonic nature of light 
quanta (photons) brings in both features, namely amplification as well 
as attenuation as the photon number is not conserved \cite{pradhan,zhang,joshi}. 
\vspace{2.0cm}   

\centerline{ {\bf Acknowledgments}}

Authors thank Professors N. Kumar and P. A. Mello for several useful 
discussions on related issues in mesoscopic systems.

\end{document}